\documentstyle[epsf]{article}
\newcommand{\gsim}{\mbox{\raisebox{-.3em}{$\stackrel{>}{\sim}$}}}
\newcommand{\lsim}{\mbox{\raisebox{-.3em}{$\stackrel{<}{\sim}$}}}

\renewcommand{\cite}[1]{\ref{#1}}

\newcommand{\half}{\frac{1}{2}}

\newcommand{\beq}{\begin{equation}}
\newcommand{\eeq}{\end{equation}}
\newcommand{\beqa}{\begin{eqnarray}}
\newcommand{\eeqa}{\end{eqnarray}}
\newcommand{\bpr}{\begin{problem}}
\newcommand{\epr}{\end{problem}}
\newcommand{\bcent}{\begin{center}}
\newcommand{\ecent}{\end{center}}
\newcommand{\bfig}{\begin{figure}}
\newcommand{\efig}{\end{figure}}
\newcommand{\bpc}{\begin{picture}}
\newcommand{\epc}{\end{picture}}

\newcommand{\nnb}{\nonumber}
\newcommand{\reflef}{(\ref}
\newcommand{\MP}{M_{\rm P}}

\begin{document}
\baselineskip = 0.6cm

\bcent
{\LARGE\bf Mass of the dilaton and the cosmological constant}\\[.4em]
Yasunori Fujii \\[.0em]
Advanced Research Institute for Science and Engineering, 
Waseda University, Shinjuku, Tokyo 169-8555, Japan
\ecent

\noindent
\bcent
{\large\bf Abstract}\\[1.8em]
\begin{minipage}{13cm}
{\small
One might raise a question if the gravitational scalar field (dilaton) mediates a finite-range force between local objects still behaving globally as being massless to implement the scenario of a decaying cosmological constant.  We offer a non-negative reply by a detailed analysis of the field-theoretical quantization procedure in relation to the observationally required suppression of the vacuum-energy as part of the contribution to the cosmological constant.
}
\end{minipage}
\ecent
\mbox{}

\section{Introduction}

The gravitational scalar field, sometimes called the dilaton of a likely origin in string theory, is a focus of attempts to understand the small but nonzero cosmological constant [\cite{r-p}].  The fast-falling potential, like the exponential or the inverse-power type, has been favored either in phenomenological analyses [\cite{peeb}] or as models of theoretical interest [\cite{YM}-\cite{Damour}].  In particular we have studied the exponential potential in the context of the scalar-tensor theory with the constant $\Lambda$  included [\cite{yfptp99},\cite{cup}].  It seems agreed that the global behavior of the {\em spatially-uniform} solution for the scalar field that allows us to implement the scenario of a decaying cosmological constant is consistent with ``masslessness" of the field, in the sense that its inverse ``mass" is as large as $t_0$, the present age of the universe, or the size of the visible universe.

Some of the authors go even further to argue that the field mediates a long-range force between macroscopic objects around us, thus is subject to the solar-system experiments, or some free-fall measurements above the detectable level [\cite{Wett2},\cite{Damour},\cite{Carroll}].  The present author, on the other hand, emphasizes that the spatially-dependent component results more likely in the local force with a finite and intermediate force-range, as also discussed in [\cite{Wett2}], not necessarily constrained by the above mentioned experiments [\cite{yfptp99},\cite{cup}].  In this article we try to reinforce the latter point of view, expressed rather vaguely in the middle of Chapter 6.4 of [\cite{cup}], by the detailed analysis of the self-energy of the quantized scalar field arising from the interaction with matter fields.

The root of the issue lies in the widely accepted view that, unlike a gauge field, the scalar field shows no immunity against acquiring self-mass due to the interaction.  We recognize, on the other hand, that the vacuum-energy in the sense of the relativistic quantum field theory predicts too much contribution to the cosmological constant, and appears to be suppressed almost to zero according to the observation.  We show how the latter fact is related to the present issue, also discussing what the theoretical formulation should be like in order to have the expected distinction between the global and local aspects.  We admit that the conclusion, partly a conjecture, is still tentative depending to some extent on the specific model chosen.  We still believe the effort to provide an important insight to the issue.

\section{Preliminaries}

We start with the 4-dimensional Lagrangian of the scalar-tensor theory with the constant $\Lambda$ included in the J(ordan) conformal frame (CF), sometimes called the string CF, or the theoretical CF:
\beq
{\cal L} = \sqrt{-g}\left( \half\xi\phi^2 R -\half\epsilon g^{\mu\nu}\partial_\mu \phi \partial_\nu \phi  -\Lambda +L_{\rm matter}  \right),
\label{md-1}
\eeq
where $\phi$ is the scalar field, while $\xi$ and $\epsilon$ are related to $\omega$ in the prototype Brans-Dicke model by $\xi^{-1} = 4|\omega|>0$ and $\epsilon =\mbox{Sign}(\omega)$.  Notice that choosing $\epsilon =-1$ implies a ghost with negative kinetic energy of $\phi$, but, due to the ``mixing" effect in the non-minimal coupling term, the first term on the right-hand side of \reflef{md-1}), the total energy remains positive as long as $\zeta^{-2}\equiv 6 + \epsilon\xi^{-1}=2( 3 +2\omega) >0$, assumed also of the order unity.  We use the (reduced) Planckian unit system with $c=\hbar =M_{\rm P} (= (8\pi G/c\hbar)^{-1/2})=1$.  Note that ${\rm GeV}= 0.41\times 10^{-18}$ while the present age of the universe $t_0\sim 14 {\rm Gy}$ is  $1.66\times 10^{60}$. To make the paper to be minimally self-contained, we summarize the formulation briefly according to [\cite{cup}], though some of the contents had been obtained in [\cite{YM}-\cite{Wett2}] in different contexts.

The cosmological constant problem has two faces [\cite{cup}]: ``Why is it so small compared with $M_{\rm P}^4$?" and ``Why is it still nonzero?"  In this note we confine ourselves to the first one, which is replied successfully by showing that the effective cosmological constant decays with the cosmic time.  Including  a constant $\Lambda$ on the right-hand side  of \reflef{md-1}), however, changes the cosmological solution without $\Lambda$ so drastically that many traditional views deviate dangerously from what the standard cosmology has told us.  In our proposed remedy [\cite{yfptp99},\cite{cup}], we allow the scalar field to enter $L_{\rm matter}$, thus violating Weak Equivalence Principle, but in such a way that its observable effects remain relatively small.   In this ``scale-invariant model," scale invariance observed by the first two terms on the right-hand side  of \reflef{md-1}) is extended to the matter part.

\begin{itemize}
\item In the prototype BD model, $L_{\rm matter}$ is assumed to be decoupled from $\phi$ in the J frame.  We revise this by introducing a term $-(1/2)f^2\phi^2 \Phi^2$, where $\Phi$ is a real scalar field as a representative of matter fields throughout this paper, not to be confused with the gravitational $\phi$, while $f$ is a dimensionless coupling constant.  No mass term of $\Phi$ is assumed at this stage. 
\item By applying a conformal transformation moving to the E(instein) CF, the above-mentioned interaction term becomes a mass term, $-(1/2)m^2\Phi_*^2$ with $m^2 = \xi^{-1}f^2$, with the scalar field left entirely decoupled from the matter field $\Phi_*$.  Scale invariance is shown to be broken {\em spontaneously} with the canonical $\sigma =\zeta^{-1}\ln (\xi^{1/2}\phi)$ as a massless Nambu-Goldstone boson.  We may accept this E frame to be a {\em physical} CF.
\item By including quantum corrections {\em among matter} fields, the $\sigma$-matter coupling re-emerges as an effect of quantum anomaly:
\beq
L' =  -\half g_\sigma \Phi_*^2 \sigma ,\quad\mbox{with}\quad g_\sigma = \zeta Q m^2 \MP^{-1},
\label{md-1-1}
\eeq
to the one-loop approximation, where $Q$ is a coefficient which depends on the non-gravitational coupling constants of the matter fields, WEP thus being violated.  The factor $\MP^{-1}$ has been re-installed as a reminder that $g_\sigma$ has mass dimension 1.  Scale invariance is now broken {\em explicitly}, leaving $\sigma$ as a pseudo NG boson, which is generically massive.  We point out that the calculation was made in the dimensional regularization method which is {\em different} from a simple cutoff procedure [\cite{tHVFM}]. 

\end{itemize}

Let us ignore the quantum effects for the moment, and re-express \reflef{md-1})  in terms of the starred quantities in the E frame:
\beq
{\cal L} = \sqrt{-g_*}\left( \half R_* -\half g_*^{\mu\nu}\partial_\mu \sigma \partial_\nu \sigma  -\Lambda e^{-4\zeta\sigma} +L_{\rm *matter}  \right).
\label{md-2}
\eeq
Note that the constant term $-\Lambda$ in \reflef{md-1}) has been transformed to an exponential potential, $V(\sigma) = \Lambda e^{-4\zeta\sigma}$.  In the Friedmann universe with $k=0$, the cosmological equations are
\beqa
3H_*^2 &=& \rho_\sigma +\rho_*, \label{md-3}\\
\ddot{\sigma} &+& 3H_* \dot{\sigma} + V'(\sigma)=0, \label{md-4}
\eeqa
where we assumed radiation dominance to simplify the equations, though the conclusion remains essentially unchanged for dust dominance as well.  We also use the over-dot for differentiation with respect to $t_*$, the cosmic time in the E frame, while $\rho_\sigma = \dot{\sigma}^2/2 +V(\sigma)$, representing ``dark energy," plays the role of an effective cosmological ``constant" $\Lambda_{\rm eff}$ in the E frame.  On the other hand, $\rho_*$ is the energy density of the matter including dark matter.

We find the solution for the spatially uniform $\sigma$:
\beq
\sigma (t_*)=\bar{\sigma} +\half\zeta^{-1} \ln t_*,
\label{md-7}
\eeq
where $\bar{\sigma}$ is a constant determined by $\Lambda e^{-4\zeta\bar{\sigma}}=(1/16)\zeta^{-2}$.  This describes a slowly-rolling-down behavior, and is an attractor solution independent of the initial values.  We also obtain 
\beqa
\rho_\sigma &=& \frac{3}{16}\zeta^{-2}t_*^{-2}, \label{md-8} \\
\rho_* &=& \frac{3}{4}\left( 1-\frac{1}{4}\zeta^{-2} \right) t_*^{-2}. \label{md-9}
\eeqa
Note that these asymptotic behaviors are independent of $\Lambda$.  Equation \reflef{md-8}) shows that $\Lambda_{\rm eff}$ falls off like $t_*^{-2}$, implementing the decaying $\Lambda$ scenario.  Note that the common ``scaling" behavior as in \reflef{md-8}) and \reflef{md-9}) provides short of understanding the ``second face," requiring a further attempt, like the ``two-scalar model" [\cite{cup}].

The potential has no minimum.  One may nevertheless define a ``mass squared" by
\beq
\mu_{b}^2 = \frac{\partial^2 V}{\partial \sigma^2} = (4\zeta)^2 V.
\label{md-10}
\eeq
By using \reflef{md-7}) we immediately find $V \sim t_*^{-2}$. Combining this with \reflef{md-10}) we conclude
\beq
\mu_{b}^2 \sim t_*^{-2},
\label{md-11}
\eeq
up to the multiplying coefficient roughly of the order unity.  At the present epoch, the mass $\mu_b$ is as small as $\sim t_{*0}^{-1}$, or the ``force-range" as large as $\sim t_{*0}$, which is the size of of the visible universe.  We may consider that $\sigma_b$ is ``massless"  in any practical situation.

The fact that the right-hand side of \reflef{md-10}) is positive implies a restoring force for any small deviation of $\sigma_b$ from the solution \reflef{md-7}).  This can be combined with the presence of the frictional force given by the second term on the left-hand side of \reflef{md-4}) supporting that this solution is stable against the small perturbation of the mass squared given by \reflef{md-11}).

\section{Background and fluctuation}

One might then be tempted to consider the exchange of this field to provide a long-range force in practice.  We point out, however, that the spatially-independent field is totally alienated from the concept of a force that attracts or repels two objects.  In order to discuss a force we must include spatially-varying portion as well.  For this purpose, we consider the decomposition:
\beq
\sigma (x)= \sigma_b(t) + \sigma_f(x).
\label{md-12}
\eeq
Substituting this into \reflef{md-2}), we would find the coupled equations for $\sigma_b$ and $\sigma_f$ for the {\em background} and the {\em fluctuating} components, respectively.  The way in which the one affects the other is found to be largely asymmetric.  Roughly speaking, $\sigma_f$ remains nearly unaffected by $\sigma_b$, while opposite in the other way.  This can be seen qualitatively by looking at the kinetic term  on the right-hand side of \reflef{md-2}), which is split into
\beq
-\half (\partial\sigma_b)^2  -(\partial\sigma_b \partial\sigma_f) -\half(\partial\sigma_f)^2.
\label{md-14}
\eeq
The last term is expected to be of the order of $\gsim \mu_f^2$, with $\mu_f$ being the self-mass of $\sigma_f$, which will be estimated to be somewhere around $\mu_f \sim m^2/\MP \sim 10^{-18}{\rm GeV}\sim 10^{-36}$ where $m\sim {\rm GeV}\sim 10^{-18}$ is a typical hadronic mass [\cite{yfnat}].  By using \reflef{md-7}), on the other hand, we expect that the first and the second terms are of the size of $t_0^{-2}$ and $t_0^{-1}\mu_f$, and are smaller than the last term by 48 and 24 orders of magnitude, respectively.   In the subsequent analyses we confine ourselves entirely to the E frame, suppressing $*$ to simplify the notation.

In order to outline a possible theoretical approach, we present the argument in two steps; (i) showing how $\sigma_f$ is quantized then acquiring mass, (ii) discussing what kind of back-reaction $\sigma_b$ can be subject to.  

\section{Quantizing  the fluctuating part}

The potential for $\sigma_f$ is given by
\beq
 -V\left( \sigma_b+\sigma_f \right)\approx -\frac{1}{16}\zeta^{-2}t_0^{-2}e^{-4\zeta\sigma_f},
\label{md-16}
\eeq
where we have used $V(\sigma_b)= (1/16)\zeta^{-2}t_0^{-2}$ for $t\approx t_0$ as derived from \reflef{md-7}).

The exponential potential allows no stationary point $dV/d\sigma_f =0$ to occur.  But this ``tilt" as well as ``curvature," both proportional to $\sim t_0^{-2}$, are again so small that any local experiments or observations, like measuring a $\sigma_f$-exchanged force between two objects, have no resolving capability to detect them.  In this approximate and realistic situation, $\sigma_f$ is well separated from $\sigma_b$, thus behaving as a {\em massless} scalar field, with
\beq
L_f= -\half(\partial\sigma_f)^2,
\label{md-17}
\eeq
which provides a classical theory based on which perturbative quantum theory around the present epoch is to be developed.

\bfig[b]
\hspace*{2.25cm}
\epsfxsize=10cm
\epsffile{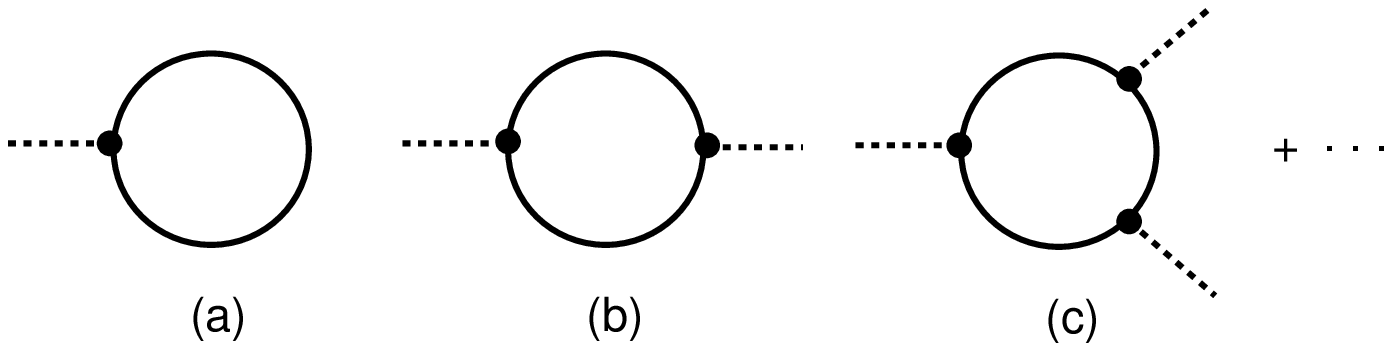}
\caption{
First three examples of the 1-loop diagrams, $\sim \sigma, \sigma^2, \sigma^3$, for the quantum corrections due to the matter coupling.  Solid and dotted lines are for $\Phi$ and $\sigma$, respectively.  }
\label{fig1}
\efig

Now according to the relativistic quantum field theory on nearly flat spacetime, the interaction \reflef{md-1-1}) with the matter field $\Phi$, which is otherwise assumed to be free, yields a tadpole:
\beq
{\cal T} = g_\sigma i(2\pi)^{-4}\int d^4k \frac{1}{k^2 + m^2},
\label{md-17-1}
\eeq
also depicted in Fig. 1(a).  The integral is divergent by power counting, but is probably cut off at some mass-scale basically of the order of magnitude not very much different from $m$, thus 
\beq
{\cal T}\sim m^4,
\label{md-17-3}
\eeq
which is $\sim 10^{-72}$, allowing a latitude of a few orders of magnitude.

We next consider the self-energy part $\Pi(p^2)$ of the quantized field $\sigma_f$ coming from the $\Phi$ loop, as shown by Fig. 1(b):
\beq
\Pi (p^2) = g_\sigma^2 {\cal J}(p^2),
\label{md-18}
\eeq
with
\beqa
 {\cal J}(p^2)&=&i (2\pi)^{-4} \int d^4k_1 \int d^4k_2 \frac{\delta^4 (k_1 +k_2 -p)}{(k_1^2 + m^2)(k_1^2 + m^2)}, \nnb\\[.5em]
&=&-\frac{1}{\pi}\int _{s_0}^{\infty} \frac{\rho(s')}{s' -s -i\epsilon}ds',
\label{md-19}
\eeqa
where $s= -p^2 = p_0^2 -\vec{p}\;^2$ with $\rho(s) = (16\pi)^{-1} \sqrt{1-s_0/s}=k/(8\pi W)$,  
where $W = \sqrt{s}$ is the total energy of the pair of $\Phi$, $k$ and $s_0 =4m^2$ are the relative momentum and the threshold energy squared, respectively.

Following the usual procedure, $\Pi$ evaluated at $p^2=-\mu_f^2$ will give a self-mass squared $\mu_f^2$, which is assumed to be positive.  It also seems reasonable to expect that $\mu_f^{-1}$ is of a macroscopic size certainly shorter than the distances relevant to the solar-system experiments [\cite{cup},\cite{yfnat}].


\section{Back-reaction to the background part}

We started by accepting the solution \reflef{md-7}) to be a classical background solution.   As explained before, however, the back-reaction from the quantized $\sigma_f$ might be serious.  One might be forced even to include ``own virtual quantum processes [\cite{BW63}]" of $\sigma_b$, despite its huge macroscopic size which may justify the classical nature.  A self-consistent way of solving the equations seems imperative, as shown by a field theoretical analysis of the non-equilibrium Bose-Einstein condensate [\cite{matsumoto}], for example.  This is a feature to be supplemented to the formal procedure described on the use of the nontrivial classical background field [\cite{tH}].

The field $\sigma_b$ thus acquires the same tadpole term as \reflef{md-17-1}) and \reflef{md-17-3}), because the coupling \reflef{md-1-1}) has been derived for the entire sum \reflef{md-12}) of $\sigma$.  The result \reflef{md-17-3}) is then added to the ``slope"
\beq
\frac{\partial V}{\partial \sigma} = -4\zeta V(\sigma_b(t_0)),
\label{md-20-2}
\eeq
which is $\sim t_0^{-2}$, as estimated in \reflef{md-10}) and \reflef{md-11}), being smaller than the ``correction"  \reflef{md-17-3}) by as much as 48 orders of magnitude.

This is a place where one might give up the whole attempt to successfully implement the scenario of a decaying $\Lambda$.  According to the conventional renormalization program,  however, the ``divergent" term is {\em replaced} by the ``observed" value.  Equivalently, one has a freedom to add a {\em counter-term} ${\cal C}_1$ to obtain the observed value ${\cal V}_1 = -4\zeta V+{\cal T} +{\cal C}_1$.  Unfortunately, no appropriate ``observation" is available for the scalar field.  We find, however, a substitute from another phenomenon of different but related kind.

We re-emphasize that $\sigma_b(t)$ is spatially constant and the energy density computed as the quantum corrections to $V(\sigma_b)$ as shown in Fig. 1, with the dotted line for the small perturbation of $\sigma_b$, is distributed uniformly in the universe, thus constituting part of the cosmological ``constant" at the epoch $t$.  Obviously this is part of the vacuum-energy without the external matter field $\Phi$.  On the other hand, the observed vacuum-energy contribution to the cosmological constant is known to be smaller than the theoretical prediction by about ``60" orders of magnitude [\cite{wnb}], which is basically the same as the number 48 derived before, in view of uncertainties in the estimates.  We now propose to use this ``suppression" as the ``observed value" for the back-reaction effect to $V(\sigma_b)$.

Toward the end of Chapter 4.4.2 of [\cite{cup}], we offered a conjecture that no discrepancy of ``60" orders of magnitude mentioned above might occur if the vacuum-energy {\em builds up gradually} starting from an infinitesimal amount in the expanding universe, unless no ingredient of the vacuum-energy is time-dependent, allowing the expression $g_{\mu\nu}\times {\rm const}$ in the {\em E-frame} Lagrangian.  This is in fact what we usually do for the matter density arriving at an asymptotic behavior $\sim t^{-2}$ like in \reflef{md-9}), with an added  reminder that we thereby must have included the vacuum-energy as well, although it could act as a cosmological ``constant" only if it falls off at least more slowly than $\sim t^{-2}$.

We note that the spatially-averaged cosmological matter density has no distinction between vacuum- and non-vacuum-energies, perhaps up to different equations of state.  In any case it seems highly unlikely that the vacuum-energy component, whether it behaves like a constant or not, grows overwhelmingly beyond the critical density at the expense of the non-vacuum energy.  The vacuum-energy  must be suppressed below the level $\sim t_0^{-2}$ for whatever the reason yet to be elaborated.  According to our proposal, the tadpole term $\sim m^4$ should be renormalized to a value $\lsim t_0^{-2}$:
\beq
{\cal V}_1 \lsim -4\zeta V,
\label{md-20-3}
\eeq  
implying that ${\cal C}_1$ is adjusted to cancel most of ${\cal T}$, probably exploiting the smallness of $t_0^{-1}$ by appealing to a cosmological argument.

A more careful analysis is needed for the self-energy part $\Pi$, because it has a spatial extension much smaller than $t_0$.  For this purpose we use the spectral representation [\cite{ptp61}] given by the second line  of \reflef{md-19}) which is convenient because the spatial Fourier transform ${\cal J} (r)$ is given in terms of the superposition of the Yukawa terms:
\beq
{\cal J}(r) =-\frac{1}{4 \pi^2}\int \frac{e^{-\kappa r}}{r}\rho(s)ds,
\label{md-21}
\eeq
where $\kappa = \sqrt{s}$.  We estimate
\beq
\Pi(r) \sim  -A\frac{g_\sigma^2}{4\pi^2} m^2 \frac{e^{-2\bar{m}_1 r}}{4\pi r},
\label{md-22}
\eeq
where $A$ is a coefficient of the order one, while $\bar{m}_1$ is somewhere between $m$ and $m_1$ below which $s$ can be approximated by $4m^2$, hence chosen roughly to be $m$.

We consider Dyson's equation:
\beq
\Delta'^{-1} = \Delta^{-1} -<\Pi>,
\label{exsp-2}
\eeq
where the zero-th order term $\Delta^{-1} = \mu_b^2 \sim t_0^{-2}$ is a constant for the spatially non-propagating field, while the last term on the right-hand side is for an ``average" $<\Pi> =  \int d\vec{r}\: \Pi (\vec{r})$. On integrating \reflef{md-22}) we find $\int d\vec{r} r^{-1}e^{- r/\lambda} \sim \lambda^2,$ where $\lambda \sim (2\bar{m}_1)^{-1} \sim m^{-1}$.  In this way we obtain
\beq
<\Pi> \sim m^6 \lambda^2 \sim m^4,
\label{exsp-11}
\eeq
which is ``large" in the same order of magnitude as in \reflef{md-17-3}).  Nearly the same argument as the tadpole term applies again.

Unlike with the tadpole, however, the function $\Pi$ depends on $p^2$.  The above-mentioned suppression is related to the global structure, and applies only to $p^2 \approx -t_0^{-2}$, leaving the value for $p^2 \approx -\mu_f^2$ obtained from the quantized component almost unaffected.  This scenario, which is yet to be justified rigorously, should make the two aspects compatible to each other; massless global field and massive local field.  One might interpret this also as a ``running" counter-term which varies toward $p^2 \approx -t_0^{-2}$ in a non-trivial manner, represented symbolically by
\beq
{\cal C}_2(-t_0^{-2}) \approx -\Pi(-t_0^{-2}),\quad \mbox{and}\quad {\cal C}_2(-\mu_f^2) \approx \Pi(-\mu_f^2).
\label{exsp-11-1}
\eeq
How good the first approximate equation is would determine how close the resulting potential is to the original one.  This requires a more precise analysis based on the standard renormalization procedure.  On the other hand, the ``observed" suppression of the vacuum energy in the cosmological constant might be so strong, as expressed by the number ``60" orders of magnitude, that the whole ``correction," including the counter terms, to $V(\sigma_b)$ can be made negligibly small.  In this simplest imaginable but still likely situation, the exponential potential remains virtually unaffected, no matter how complicated the detailed renormalization procedure might be.

Similar analyses on the higher-order terms of $\sigma_b$, like the one illustrated by Fig. 1(c), should be subject to the same kind of suppression required for all the terms, in order to maintain the exponential potential.  This necessity on the finite-part renormalization might appear too artificial, but is supported by the suppression of all kinds of the vacuum-energy.  It is rather natural after all to expect that the influence of the unique nature of $\sigma_b$ which extends globally to the whole universe overrides consequences of the local physics of the quantized component as far as the global aspect is concerned.

\section{Concluding remark}

We have shown that globally massless and locally massive behaviors of a single field dilaton can be consistent with each other.  Although the conclusion is not unique, this feature is related closely to another theoretical question why the vacuum-energy expected naturally from the quantum field theory in Minkowskian spacetime predicts too much excess compared with the observation.  We re-iterate that we have only outlined what the theory should be like.  Solving the equations for $\sigma_b$ and $\sigma_f$ more rigorously should be made only if we establish a detailed mechanism to suppress the vacuum-energy.  This will be a key to understand the whole problem of the cosmological constant including the scenario of a decaying $\Lambda$ for the component of the primordial origin. \\[.4em]

I thank Thibault Damour whose comment on the occasion of JENAM 2002 in Porto-Portugal motivated this work.  Thanks are also due to Y. Chikashige, R. Fukuda, S. Ichinose, S. Kamefuchi, K. Kawarabayashi, K. Maeda, H. Matsumoto  and M. Omote for their enlightening discussions.

\newpage
\mbox{}\\
\noindent
{\Large\bf References}

\begin{enumerate}
\item\label{r-p}A.G. Riess {\em et al.}, Astgron. J. {\bf 116}, 1009 (1998); S. Perlmutter {\em et al}., Nature, {\bf 391}, 51 (1998). 
\item\label{peeb}B. Ratra and P.J.E. Peebles, Phys. Rev. {\bf D37}, 3406 (1988); R.R. Caldwell, R. Dave and P.J. Steinhardt,   Phy. Rev. 
Lett. {\bf 80}, 1582 (1998); L. Wang, R.R. Caldwell, J.P. Ostriker,
P.J. Steinhardt,  Astrophys.J. {\bf 530}, 17 (2000).
\item\label{YM}J. Yokoyama and K. Maeda, Phys. Lett. {\bf B207}, 31 (1988).
\item\label{Wett1}C. Wetterich, Nucl. Phys. {\bf B302}, 645 (1988).
\item\label{Wett2}C. Wetterich, Nucl. Phys. {\bf B302}, 668 (1988).
 
\item\label{Damour}T. Damour, F. Piazza and G. Veneziano, Phys. Rev. Lett. {\bf  89}, 081601 (2002): gr-qc/0204094, Phys. Rev. {\bf D66}, 046007 (2002):  hep-th/0205111.
\item\label{yfptp99}Y. Fujii, Prog. Theor. Phys. {\bf 99}, 599 (1998).
\item\label{cup}Y. Fujii and K. Maeda, {\sl The scalar-tensor theory of gravitation,} Cambridge University Press, (2003).
\item\label{Carroll}S.M. Carroll, Phys. Rev. Lett. {\bf 81}, 3067 (1998): astro-ph/9806099; T. Chiba, Phys. Rev. {\bf D60}, 083508 (1999): gr-qc/9903094; N. Bartolo and M. Pietroni, Phys. Rev. {\bf D61}, 203518 (2000); hep-ph/9908521.
\item\label{tHVFM}G. 'tHooft and M. Veltman, Nucl. Phys. {\bf B44}, 189 (1972); Y. Fujii and K. Mima, Prog. Theor. Phys.  {\bf 58}, 991 (1977).
\item\label{yfnat}Y. Fujii, Nature Phys. Sci. {\bf 234}, 5 (1971), Phys. Rev. {\bf D9}, 874 (1974).

\item\label{BW63}B. DeWitt, Dynamical theory of groups and fields, in {\sl Relativity, groups and topology}, Gordon and Breach, New York, 1963.  See page 787.
\item\label{matsumoto}H. Matsumoto and S. Sakamoto, Prog. Theor. Phys. {\bf 105}, 573 (2001).
\item\label{tH}B. DeWitt, Phys. Rev. {\bf 162}, 1195 (1967); G. 'tHooft, Nucl. Phys. {\bf B62}, 444 (1973).
\item\label{wnb}S. Weinberg, Rev. Mod. Phys. {\bf 61}, 1 (1989).
\item\label{ptp61}Y. Fujii, Prog. Theor. Phys. {\bf 26}, 391 (1961).

\end{enumerate}

\end{document}